\documentclass[12pt]{article}
\usepackage{epsfig}
\usepackage{cite}
\newcommand{\ra}{\rightarrow}
\newcommand{\be}{\begin{equation}}
\newcommand{\ee}{\end{equation}}
\newcommand{\ba}{\begin{array}}
\newcommand{\ea}{\end{array}}
\newcommand{\pa}{\partial}
\newcommand{\bi}{\bibitem}

\begin{document}
\begin{center}
\ \\
{\bf\large Glueball masses and Regge trajectories for
the QCD-inspired potential}\\
\date{ }
\vskip 7mm
M.N. Sergeenko\\
\vskip 3mm
{\it Institute of Physics, Belarus National Academy of Sciences,\\
 68 Nezavisimosti Ave., Minsk, 220072, Belarus}\\
({\small E-mail: msergeen@mail.com})
\end{center}

\vskip 8mm \centerline{\bf Abstract}{\small
Bound state of two massive constituent gluons is studied in
the potential approach. Relativistic quasi-classical wave equation with
the QCD-inspired scalar potential is solved by the quasi-classical
method in the complex plane. Glueball masses are calculated with the
help of the universal mass formula. The hadron Regge trajectories are
given by the complex non-linear function in the whole region of the
invariant variable $t$. The Chew-Frautschi plot of the leading glueball
trajectory, $\alpha_P(t)$, has the properties of the $t$-channel Pomeron,
which is dual to the glueball states in the $s$ channel. The imaginary
part of the Pomeron is also calculated.}

\vskip 5mm
\noindent Pacs: 12.39.Mk; 12.39.Pn; 12.40.Nn; 12.40.Yx

\section{Introduction}
Quantum Chromo Dynamics allows the existence of purely gluonic bound
states, glueballs. These are particles whose valence degrees of
freedom are gluons where the gauge field plays a more important dynamical
role than in the standard hadrons. Study of glueballs is a good test of
our understanding of the non-perturbative (NP) aspects of QCD, but no
firm experimental discovery of such gluonic states has been obtained yet.
A comprehensive review devoted to the glueballs was given in
\cite{MaKoVe,BBMS,KleZa}.

An important theoretical achievement in this field has been the
computation of the glueball masses in lattice QCD \cite{LattGl}. Lattice
QCD has been able to compute the low-lying glueball spectrum with a
good accuracy. This predictions for glueballs are now fairly stable,
at least when virtual quarks are neglected. Several candidates for the
low mass glueballs with quantum numbers $J^{PC}=0^{++}$, $2^{++}$,
$0^{-+}$ and $1^{--}$ are under discussion. It is difficult to single
out which states of the hadronic spectrum are glueballs because we lack
the necessary knowledge to determine their decay properties.

Many theoretical approaches contain basic model assumptions which
are difficult to prove starting from the QCD Lagrangian. Complete
understanding of glueballs includes such theoretical treatments as
lattice QCD, constituent models, AdS/QCD methods, and QCD sum rules
\cite{MaKoVe,KleZa,MaBuSe}.
Glueballs have been studied by using effective approach like Coulomb
gauge QCD and potential model \cite{MaSeSi,BraSe,KaiSi}. The potential
model is very successful to describe bound states of quarks. It is also
a possible approach to study glueballs \cite{MaBuSe,BraSe}. Recent
results in the physics of glueballs with the aim set on phenomenology
and the possibility of finding them in conventional hadronic experiments
have been reviewed in \cite{MaBuSe}

A possible way to handle glueballs is to consider massive quasi-gluons
interacting via a QCD inspired dynamics. The gluons are massless
to all orders in perturbation theory, but NP effects like confinement,
and their self-interactions, can be described by a constituent gluon mass.
Another definition of the gluon mass was considered in \cite{Corn}, where
a realistic QCD motivated gluon propagator was obtained from approximate
solution of the Dyson-Schwinger equation. The dynamical mass of gluon
is defined by the position of the pole of the dressed gluon propagator.

One of open topics in hadron physics is the Pomeron. What is the Pomeron?
We know the Pomeron as the highest-lying Regge (Pomeranchuk) trajectory
($P$ trajectory) \cite{DoDoLa}. In the Regge pole theory, the leading
Regge trajectories give the main contribution into the scattering
amplitude. In the many high energy reactions with small four-momentum
transfer, the soft $P$ exchange gives the dominant contribution in cross
sections \cite{DLPW}. Next question is: what is the relation between
glueballs and the Pomeron? There has been a long-standing speculation
that the physical particles on the $P$ trajectory might be glueballs
\cite{KleZa,MaBuSe,MaSeSi}.

The relation between glueballs and the Pomeron has been investigated by
many authors. Usually, the $P$ trajectory is considered to be a
linear function. However, recent small $-t$ ZEUS and H1 data for
exclusive $\rho$ and $\phi$ photoproduction \cite{ZEUS,Ast,HERA} point
out that the $P$ trajectory is rather non-linear. The data have
been explained by adding in a flavor-blind hard Pomeron contribution,
whose magnitude is calculated from the data for exclusive $J/\Psi$
photoproduction \cite{DL,DLZ}. The ZEUS, H1 as well as CDF data on
$p\bar p$ elastic scattering data have also been analyzed by using
the non-linear $P$ trajectory \cite{FiJPP,Godiz,GodPet,SeEPL}.

In this work we consider two-gluon glueballs as the excited states of
purely gluonic bound states of massive gluons. We accept the potential
model, which is so successful to describe bound states of quarks;
it is also a possible approach to study glueballs \cite{BraSe,HoLW}.
To describe the two-body system we use the relativistic quasi-classical
(QC) wave equation with the QCD-inspired scalar potential, in which
the strong coupling is the coordinate dependent, i.e.,
$\alpha_s=\alpha_s(r)$. We obtain two exact solutions of the relativistic
QC wave equation for two components of the potential, the short-distance
coulombic term and long-distance linear one, separately. Using the
interpolation procedure, we join these two solutions and obtain an
interpolating mass formula, for the bound system. Using this universal
mass formula, which is good to describe the mass spectra of both light
and heavy quarkonia, we calculate glueball masses and reconstruct the
saturating $P$ trajectory.

This work is not a comprehensive investigation of the glueball
spectroscopy. We concentrate ourself on the leading $S_z=2$ gluonium
states and take a picture where the $t$-channel Pomeron is
dual to glueballs in the $s$ channel. We obtain an analytic expression
for the $P$ trajectory, $\alpha_P(t)$, in the whole region of
the Mandelstam invariant variable $t$. The trajectory is a complex
non-linear function, the real part of which corresponds to the soft
Pomeron in agreement with the recent HERA data and is the saturating
trajectory. The imaginary component of the $P$ trajectory is also
calculated.

\section{Glueballs and the Pomeron}

Glueballs were suggested theoretically in \cite{Min} and then have
been extensively studied in the framework of different approaches
\cite{KleZa,MaBuSe,HoLW,Di}. These objects have not been an easy
subject to study due to the lack of phenomenological support. Much
debates have been associated with their properties. The main
achievement of these debates is the understanding of the deep relation
between the properties of the glueball states and the structure of
the QCD vacuum. The basic idea is that the vacuum is filled with
$J^{PC}=0^{++}$ transverse electric glueballs which form a
negative energy condensate \cite{DonJo}.

Glueballs are bosons made only from the gluonic field; these are
quarkless hadrons. They can be classed as mesons, because they are
hadrons and carry zero baryon number. Glueballs must be flavor singlets,
i.e., have vanishing isospin ($I=0$) and strangeness. Like all particle
states, they are specified by the quantum numbers which label
representations of the Poincare symmetry, i.e., $J^{PC}$ and by the mass.
They have the same quantum numbers as isospin $0$ mesons and their
decays in conventional hadrons violate the Okubo-Zweig-Iizuka rule.

Typically, every quark model meson comes in SU(3) flavor nonets $-$ an
octet and a flavor singlet. A glueball shows up as an extra
(supernumerary) particle outside the nonet. In spite of such seemingly
simple counting, the assignment of any given state as a glueball
remains tentative even today. In a strongly coupled theory there is
nothing to stop them mixing with the 'quark-based' states. These
quarkless states are extremely difficult to identify in particle
accelerators, because they mix with ordinary meson states. Spectrum
of pure SU(3) Yang-Mills states has been extracted using computerized
lattice calculations \cite{LattGl}.

Pure gauge QCD has been investigated by lattice QCD for many years.
This led to a well established glueball spectrum below 4 GeV
\cite{LattGl}. Lattice QCD has been able to compute the low-lying
glueball spectrum with a good accuracy. The data shows five isoscalar
resonances — $f_0(600)$, $f_0(980)$, $f_0(1370)$, $f_0(1500)$ and
$f_0(1710)$. Of these the $f_0(600)$ is usually identified with the
$\sigma$ of chiral models. The decays and production of $f_0(1710)$
give strong evidence that it is also a quarkless meson.

At the present time, the three states $2^{++}$, $0^{++}$, and $0^{-+}$
are chosen as possible experimental glueball candidates \cite{ZouBu}
because they are computed with relatively small errors in lattice
calculations \cite{MornChen}. Some experimental glueball candidates
are currently known, such as the $f_0(980)$, $f_0(1500)$ and $f_0(1710)$,
but no definitive conclusions can be drawn concerning the nature of
these states. We have got no model-independent theoretical knowledge
of these hadrons. Major new experimental effort forthcoming at
Jefferson Lab.

The modern development in glueball spectroscopy from various
perspectives has been discussed in \cite{MaKoVe,MaBuSe,SeEPL}. All
these investigations support the use of an effective
gluon mass to describe the glueball dynamics of QCD. If a valence gluon
is a priori massive, then it is a spin-1 particle. Such two gluons
with total spin $S=2$ in glueballs reproduce properly the lattice QCD
spectrum for $C=+$ states \cite{LattGl} and may have five $2S+1$ spin
states: $S_z=+2,+1,0,-1,-2$. Highest glueball trajectory with $S_z=+2$ is
expected to be the Pomeron. There are works, arguing that a valence
gluon is a massless particle, which gains a constituent mass $\mu_g$,
either constant, or state-dependent \cite{BraSe,KaiSi,Corn}.

There are other definitions of the gluon mass \cite{MaKoVe,BBMS,KleZa},
which we do not discuss here. All these arguments support the use of
an effective gluon mass to describe the dynamics of QCD. It is therefore
possible to envisage an approach to bound states made of constituent
massive gluons. We consider here the simplest case of two-gluon
glueballs, since they have always a positive conjugation charge.

An open topic in hadron physics is the relation between glueballs
and the Pomeron. In gauge theories with string-theoretical dual
descriptions, the Pomeron emerges unambiguously. In the QCD framework
the Pomeron can be understood as the exchange of at least two gluons
in a color singlet state \cite{LowN}. The pQCD approach to the
Pomeron, the Balitski\v{i}-Fadin-Kuraev-Lipatov (BFKL) Pomeron, has
been discussed in \cite{BFKL}. The Pomeron can also be associated
with a reggeized massive graviton \cite{SofHardP}.

The Pomeron is the vacuum exchange contribution to scattering at
high energies at leading order in $1/N_c$ expansion. It is the
highest-lying Regge trajectory. In the many high energy reactions
with small four-momentum transfer the $P$ exchange gives the dominant
contribution \cite{DLPW}. The classic soft Pomeron is constructed
from multi-peripheral hadronic exchanges. It is usually believed that
the soft $P$ trajectory is a linear function,
\be
\alpha_P(t)=\alpha_P(0)+\alpha_P^\prime(0)t,
\label{alPt} \ee
where the intercept $\alpha_P(0)=1$ and the slope
$\alpha_P^\prime(0)=0.25$ (GeV/c)$^{-2}$. These fundamental
parameters are very important in high-energy hadron physics. Usually,
they are determined from experiment in hadron-hadron collisions.

To explain the rising hadronic cross sections at high energies,
the classic soft Pomeron was replaced by a soft supercritical
Pomeron with an intercept $\alpha_P(0)\simeq 1.08$. The approximate
linearity (\ref{alPt}) is true only in a small $-t$ region. The ZEUS,
H1 as well as CDF data on $p\bar p$ elastic scattering data have also
been analyzed by using the non-linear $P$ trajectory \cite{FiJPP}.
Important theoretical results have been obtained in \cite{ErSch}.
The results imply that the effective $P$ trajectory flattens for
$-t > 1$ (GeV/c)$^2$ that is evidence for the onset of the perturbative
2-gluon Pomeron. These results may shed some light on the
self-consistency of recent measurements of hard-diffractive jet
production cross sections in the UA8, CDF and HERA experiments.

The issue of soft and hard Pomerons has been discussed extensively
in the literature \cite{DoDoLa,BFKL,DuKaSi}. Both the IR (soft)
Pomeron and the UV (BFKL) Pomeron are dealt in a unified single step.
On the basis of gauge/string duality, the authors describe
simultaneously both the BFKL regime and the classic Regge regime
\cite{BFKL,DuKaSi}. The problem was reduced to finding the spectrum of
a single $j$-plane Schr\"odinger operator. The results agreed with
expectations for the BFKL Pomeron at negative $t$, and with the
expected glueball spectrum at positive $t$, but provide a framework
in which they are unified.

A model for the Pomeron has been put forward by Landshoff and
Nachtmann where it is evidenced the importance of the QCD NP
vacuum \cite{DoDoLa}. The current data is compatible with a smooth
transition from a soft to a hard Pomeron contribution which can
account for the rise of $\sigma_{tot}$ with $s$. If soft and BFKL
Pomeron have a common origin, the discontinuity across the cut in
the $\alpha_P(t)$ plane must have a strong $t$ dependence that points
out on non-linearity of the $P$ trajectory \cite{Bjor}.

On the theoretical front, Tang \cite{TangNo} used perturbative QCD (pQCD)
to show that Regge trajectories are non-linear by studying high energy
elastic scattering with mesonic exchange in the case of both fixed and
running coupling constants. On the experimental side, Brandt {\it et al.}
\cite{BraEr} affirmed the existence of non-linear $P$ trajectories
from the data analysis of the UA8 and ISR experiments at CERN.
They published a parametrization of $P$ trajectories containing
a quadratic term,
\be
\alpha_P(t)=1.10 +0.25t +\alpha_P^{\prime\prime}(0)t^2,
\label{alPt2} \ee
where $\alpha_P^{\prime\prime}(0)$ is a constant, which are found from
the Pomeron data fit.

Burakovsky {\it et al.} \cite{Burak} presented a phenomenological string
model for logarithmic and square root Regge trajectories. They applied a
phenomenological approach based on nonlinear Regge trajectories to glueball
states. The parameters, i.e., intercept and threshold, or trajectory
termination point beyond which no bound states should exist, were determined
from Pomeron (scattering) data. They predicted masses of glueballs on the
tensor trajectory. The approach was applied to available quenched lattice
data and found a discrepancy between the lattice based thresholds and
the Pomeron threshold that was extracted from data.

Linear trajectories are, in fact, disfavored by various experimental data.
For more details see discussions in \cite{Burak}. The square root form of
the trajectory with the parameters fitted to scattering data alone gives
the same mass predictions as the fit to both the scattering data and the
tensor glueball mass, but with larger errors. Using the fit, the authors
obtained the following predictions for excited glueball masses:
$M(2^{++})=2.38\pm 0.12$ GeV, $M(4^{++})=4.21\pm 0.21$ GeV, and
$M(6^{++})=5.41\pm 0.28$ GeV with the same central value obtained from purely
scattering Pomeron data.

Glueballs in full QCD are very complicated systems. The lightest glueballs
with positive charge conjugation $C=+$ can be successfully modeled by a
two-gluon system (gluonium) in the framework of the potential approach
\cite{MaBuSe}. Gluonium leading states and their connection with the Pomeron
have been studied in our ref. \cite{SeEPL}. We modeled glueballs to be
bound states of two constituent massive gluons interacting by the Cornell
potential and shown a good agreement both with the lattice calculations in
bound state region and the scattering data for the Pomeron.

The funnel-shaped Cornell potential is fixed in an extremely simple manner
in terms of very small number of parameters. In pQCD, as in QED the
essential interaction at small distances is instantaneous coulombic
one-gluon exchange (OGE); in QCD, it is $qq$, $qg$, or $gg$ Coulomb
scattering \cite{Bjor}. Therefore, one expects from OGE a Coulomb-like
contribution to the potential, i.e., $V_S(r)\propto-\alpha_s/r$ at $r\ra 0$.

For large distances, in order to be able to describe confinement, the
potential has to rise to infinity. From lattice-gauge-theory computations
\cite{LattV} follows that this rise is an approximately
linear, i.e., $V_L(r)\simeq\sigma r+$const for large $r$, where
$\sigma\simeq 0.15$\,GeV$^2$ is the string tension. These two
contributions by simple summation lead to the famous funnel-type (Cornell)
quark-antiquark potential \cite{LattV,Eich,QuRos},
\be
V(r)=V_S(r)+V_L(r)=-\frac 43\frac{\alpha_s}r +\sigma r;
\label{Vcor}\ee
its parameters are directly related to basic physical quantities noted
above. All phenomenologically acceptable QCD-inspired potentials are only
variations around this potential.

As for gluonium, the situation is very similar. The potential of $gg$
interaction has a similar form, but different parameters. A new method
called the Vacuum Correlator Model (VCM) has been used in \cite{KaiSi}.
In this model all NP and perturbative dynamics of quarks and gluons is
universally described by lowest cumulants, i.e., gauge invariant
correlators of the type
$\langle F_{\mu\nu}(x_1)\ldots F_{\lambda\sigma}(x_{\nu})\rangle$.
In the adjoined and fundamental representations, the final form
of interaction of two massive gluons is the funnel-type potential
of the form (\ref{Vcor}) \cite{KaiSi}:
\be
V(r)=-\frac{\alpha_a}r+\sigma_a r+C_0 \label{Vadj}, \ee
where $\alpha_a\equiv\alpha^{adj}=3\alpha_s^{fund}$,
$\sigma_a\equiv\sigma^{adj}$=$(9/4)\sigma^{fund}$; here
$\alpha_s^{fund}$ is the light quarks strong coupling,
$\sigma^{fund}\equiv\sigma\simeq 0.15$\,GeV$^2$ is the string
tension. The potential (\ref{Vadj}) was used in \cite{SeEPL} to
calculate glueball masses and the $P$ trajectory.

It is hard to find an analytic solution of the wave equation for
the potential (\ref{Vadj}). However, joining two exact solutions
obtained separately for the short-distance coulombic, $V_S(r)$,
and long-distance linear part, $V_L(r)$, of the potential, with
the help of the two-point Pad\'e approximant we obtained
the interpolating mass formula \cite{SeEPL,SeZ,SeA},
\be
M_n^2=8\tilde\sigma\left(2n_r+J +\frac 32-\tilde\alpha\right)
+4m^2\left[-\frac{\tilde\alpha^2}{(n_r+J+1)^2}+1\right],
\label{Eint} \ee
where parameters depend on the system ($q\bar q$ or $gg$):
$\tilde\alpha=(4/3)\alpha_s^{fund}$ ($q\bar q$), $3\alpha_s^{fund}$
($gg$) and $\tilde\sigma=\sigma^{fund}$ ($q\bar q$),
$(9/4)\sigma^{fund}$ ($gg$). The simple mass formula (\ref{Eint})
describes equally well the mass spectra of all $q\bar{q}$ and
$Q\bar Q$ mesons ranging from the $u\bar d$ ($d\bar d$, $u\bar u$,
$s\bar s$) states up to the heaviest known $b\bar b$ systems
\cite{SeZ,SeA}. This same formula has been used to calculate
the glueball masses as well \cite{SeEPL}.

Regge trajectories are usually assumed to be linear in $t$, but
there are both phenomenological and theoretical arguments supporting
the idea of non-linear trajectories \cite{Burak}. Inverting
(\ref{Eint}), we obtained the cubic equation for the angular
momentum $J$ and, therefore, the analytic dependence $J(M_n^2)$ for
Regge trajectories including the $P$ trajectory, $\alpha_P(t)$, in
the whole region of the invariant variable $t$ \cite{SeEPL,SeZ,SeA}.
\begin{figure}[htb]
\begin{center}
\includegraphics[scale=2]{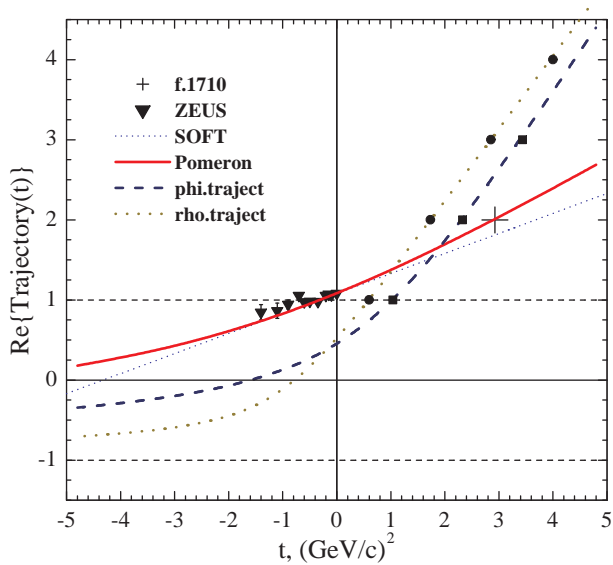}
\caption{{\small The Chew-Frautschi plots of the leading $\rho$,
$\phi$ and $P$ complex Regge trajectories calculated from the mass
formula (\ref{Eint}) with the parameters found from the fit of combined
HERA $\rho$ and $\phi$ data (triangles) \protect\cite{ZEUS,Ast,HERA},
and $2^{++}$ glueball candidate $f_0(1710)$ (cross) \protect\cite{Kirk}.
Solid curve is the $P$ trajectory given by (\protect\ref{Eint}),
dotted and dashed lines show the leading $\rho$ and $\phi$ saturating
Regge trajectories.}}\label{Fig1}\end{center}
\end{figure}
The ``saturating'' Regge trajectories in fig.\,\ref{Fig1} were applied
with success to the photoproduction of vector mesons that provide an
excellent simultaneous description of the high and low $-t$ behavior
of the $\gamma p\ra pp$, $\omega$, $\phi$ cross sections, given an
appropriate choice of the relevant coupling constants (JML-model)
\cite{CLAS,CLAS1,JLAB}. As was explained in \cite{CollKear,GuiagVan}
the hard-scattering mechanism is incorporated in an effective way
by using the ``saturated'' Regge trajectories that are independent
of $t$ at large momentum transfers \cite{SeEPL,SeZ,SeA}.

Saturating trajectories have a close phenomenological connection
to the quark-antiquark interaction which governs the mesonic
structure \cite{SeEPL,SeZ,SeA}. They provide an effective way
to implement gluon exchange between the quarks forming the exchanged
meson \cite{Lag04,Lag02} and lead to the asymptotic quark counting
rules \cite{BroFarr} that, model independently, determine the energy
behavior of the cross section at large $-t$. This approach was
successfully adopted to explain the large momentum transfer
hadron-hadron interactions, as well as several photon-induced
reactions \cite{GuiagVan}. The pion saturating trajectory
($\alpha_\pi^{sat}(t)=-1$ when $t\ra-\infty$) is in a form that
reproduces the $\gamma p\ra n\pi^+$ reaction around
$\theta_\omega^*=90$ \cite{GuiagVan}.

A fair agreement with the experiments is achieved when saturating
Regge trajectories \cite{SeEPL,SeZ,SeA} are used for the propagators
of the various exchanged mesons. This is an economical way to deal
with hard scattering mechanisms since the saturation of the Regge
trajectories (approaching $-1$ when $-t\ra\infty$) is closely related
to the OGE interaction between quarks \cite{SeZ,SeA}. The $\omega$
meson production channel is particularly instructive in this respect
since pion exchange dominates the cross section \cite{JLAB}.
\begin{figure}[htb]
\begin{center}
\includegraphics[scale=2]{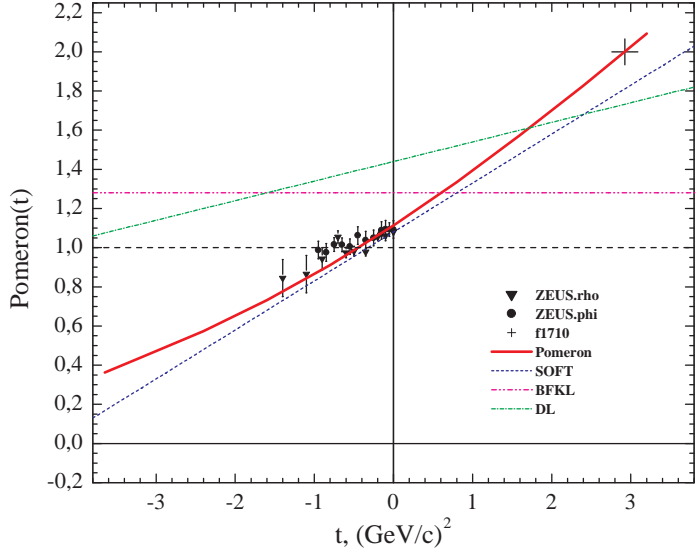}
\caption{{\small The vacuum effective $J=l+2$ Regge trajectory.
Solid curve is the $P$ trajectory obtained from (\protect\ref{Eint})
with the parameters found from the fit to combined HERA $\rho$ (triangles)
and $\phi$ (circles) data \protect\cite{ZEUS,Ast,HERA}, and $2^{++}$
glueball candidate $f_0(1710)$ (cross) \protect\cite{Kirk}. Other lines
show the classic ``soft'', BFKL, and Donnachie-Landshoff ``hard''
Pomerons.}}\label{Fig2}
\end{center}
\end{figure}
This same technics and the mass formula (\ref{Eint}) have been used
in \cite{SeEPL} to reconstruct the saturating $P$ trajectory,
$\alpha_P(t)$, in the whole region of $t$. Our $P$ trajectory and its
comparison with known others are shown in fig.\,\ref{Fig2}. The $P$
trajectory in figs.\,\ref{Fig1} and \ref{Fig2} corresponds to the soft
Pomeron and is in agreement with the HERA data \cite{ZEUS,Ast,HERA};
it is the saturating Regge trajectory, i.e., $\alpha_P(t\ra-\infty)\ra -1$.

However, the strong coupling $\alpha_s$ in the Cornell potential above
and in the mass formula (\ref{Eint}) is a constant
value (free parameter). As we know, the strong coupling in QCD is the
function of $q^2=t$, i.e., $\alpha_s=\alpha_s(q^2)$ is the running
strong coupling. Below, we introduce the dependence $\alpha_s(r)$, and
obtain similar mass formula and Regge trajectories with the use of
the QCD-inspired potential.

\section{The QCD-inspired potential}

The strong coupling in QCD is a function of the squared four-momentum
transfer, $t$: $\alpha_s(q^2=t)$, or $\alpha_s(r)$ in the coordinate space.
A more accurate calculation of hadronic masses and their trajectories
requires the accounting for the dependence $\alpha_s(r)$ in the potential
of interaction.

To find the dependence $\alpha_s(r)$ let us consider the concept of
dynamically generated gluon mass, which arises from an analysis of the
gluon Dyson-Schwinger (DS) equations \cite{Corn}. The infinite set of
couple DS equations cannot be resolved analytically. One must resort to a
truncation scheme. Cornwall found a gauge-invariant procedure to deal
with these equations \cite{Corn}.

An approximate resolution of the DS equations was obtained in the Feynman
gauge. In Euclidean space, this solution is given by
$D_{\mu\nu}=-ig_{\mu\nu}D(q^2)$, where,
\be
\alpha_0 D(q^2)=\frac{\alpha_s(q^2)}{q^2+\mu^2(q^2)}, \label{Corn}\ee
and
\be
\alpha_s(q^2)\equiv\frac{g^2(q^2)}{4\pi}=
\frac 1{b_0\ln\{[q^2+4\mu^2(q^2)]/\Lambda^2\}} \label{alfq}\ee
with the momentum dependent dynamical mass given by
\be
\mu^2(q^2)=\mu_g^2\left[\frac{\ln[(q^2+4\mu_g^2)/\Lambda^2]}
{\ln(4\mu_g^2/\Lambda^2)} \right]^{-\frac{12}{11}}. \label{mass}
\ee
Here in eqs. (\ref{Corn})-(\ref{mass}) $b_0=(33-2n_f)/(12\pi)$, $n_f$ is
number of flavors, $\mu_g=\mu(0)$, $\Lambda$ is the QCD dimensional parameter;
typical values are: $\mu_g=500 \pm 200$ MeV and $\Lambda=300$ MeV \cite{Corn}.

This solution contains a dynamically generated gluon mass (\ref{mass}) and
is another NP approach which has led to a very appealing physical
picture establishing that the QCD running coupling freezes in the NP regime.
Expression (\ref{alfq}) is considered to be the QCD running coupling in
momentum representation; it is frozen in the NP regime ($q^2\ra 0$),
\be
\alpha_0\equiv\alpha_s(0)=\frac 1{2b_0\ln(2\mu_g/\Lambda)}, \label{alf0}\ee
because of the presence of the dynamical gluon mass the strong effective
charge, $g(q^2)$, extracted from these solutions freezes at a finite value,
giving rise to an infrared fixed point for QCD \cite{Corn}. The gluon mass
generation is a purely NP effect associated with the existence of infrared
finite solutions for the gluon propagator.

Solution (\ref{Corn}) is valid only for $\mu_g>\Lambda/2$. An important
feature of the propagator (\ref{Corn}) is that it incorporates the correct
ultraviolet behavior, i.e., asymptotically obeys the renormalization
group equation. This means that the gluon propagator (\ref{Corn})
asymptotically at large $q^2$ takes the usual form, i.e., $D(q^2)\propto 1/q^2$,
because $m^2(q^2)\ra 0$ at $q^2\ra\infty$ and is valid for the entire range
of momentum. The gluon is massless at the level of the fundamental QCD
Lagrangian, and remains massless to all order in pQCD.
The NP QCD dynamics generates an effective, momentum-dependent mass, without
affecting the local SU(3)c invariance, which remains intact \cite{Corn}.

According to Low and Nussinov the Pomeron is modeled as the exchange of two
gluons \cite{LowN}. In \cite{SePRD} we modeled the $P$ exchange by two
NP gluons as suggested by Landshoff and Nachtmann \cite{LaNa}. We dealt with
the Cornwall's propagator in the context of the Landshoff-Nachtmann model
and extracted the ``pure'' NP propagator. We have shown that the last one
in combination with the multi-Pomeron asymptotic of the Quark-Gluon String
Model (QGSM) \cite{QGSM,LySeR,LySe}, results in a good description of soft
and hard distributions of secondary hadrons in a wide energy range.

Some consequences of the Cornwall's solution for the gluon propagator
associated to the static interaction were investigated in \cite{GonMaVe}.
The OGE static potential derived from the DS equations (DS potential)
was calculated numerically and compared to phenomenological potentials whose
shape has been inspired by lattice computations. Application of this DS
potential and comparison with some others to the description of quarkonia
was considered.

The strong coupling $\alpha_s$ in the Cornell potential (\ref{Vcor}),
(\ref{Vadj}), is a free parameter. This potential can be modified by
introducing the $\alpha_s(r)$-dependence, which is unknown. However,
using the mnemonic rule, $q^2\ra 1/r^2$, from (\ref{alfq}) one can
write an ansatz, for the strong running coupling in the coordinate
space as follows:
\be
\alpha_s(r)=\frac 1{b_0\ln[1/(\Lambda r)^2+(2\mu_g/\Lambda)^2]}
\label{alfr}. \ee
The running coupling (\ref{alfr}) conserves the basic properties of the
one (\ref{alfq}) in the momentum representation: $\alpha_s(r\ra 0)=0$
($q^2\ra\infty$) and $\alpha_s(r\ra\infty)=\alpha_0$ ($q^2\ra 0$).
We see, that the running coupling (\ref{alfr}) is frozen in the NP
regime ($r\ra\infty$) and is in agreement with the asymptotical freedom
properties [$\alpha_s(r\ra 0)\ra 0$].

Thus, with the help of (\ref{alfr}), we come to the following potential of
interaction:
\be
V(r)=-\frac{\tilde\alpha(r)}r+\tilde\sigma r, \label{Valr}\ee
where $\tilde\alpha(r)=k\alpha_s(r)$, $k=4/3$ ($q\bar q$ systems),
$k=3$ ($gg$ system) and $\tilde\sigma$ as in (\ref{Eint}). The
spin-dependent corrections to the potential (\ref{Valr}) can also be
derived from lattice QCD, but we do not consider them here.

In hadron physics, the nature of the potential is very important.
There are normalizable solutions for scalarlike potentials, but
not for vectorlike \cite{Su}. No any problems arise and no any
difficulties encountered with the numerical solution if the confining
potential is purely {\em scalarlike}. The effective interaction has
to be Lorentz-scalar in order to confine quarks and gluons \cite{Su}.

\section{Solution of the QC wave equation}

It is hard to find an analytic solution of known relativistic wave
equations for the potential (\ref{Valr}), that does not allow us
to get an analytic dependence $E^2(n_r,l)$. This aim can be achieved
with the use of the QC wave equation \cite{SeRelQC,SePRA}, which for
two bound particles of equal masses in the rest frame is
\be
\left[\frac{\pa^2}{\pa r^2} +\frac 1{r^2}\left(\frac{\pa^2}{\pa\theta^2}
+\frac 1{\sin^2\theta}\frac{\pa^2}{\pa\varphi^2}\right)
+\frac{E^2}4-\left(m+V\right)^2\right]\tilde\psi(\vec r)=0. \label{RelQC}\ee
This is the second-order differential equation of the Schr\"odinger
type in canonical form. Important feature of this equation is that,
for two and more turning-point problems, it can be solved exactly
by the conventional WKB method \cite{SeRelQC,SePRA,An3D}.

Appropriate solution method of the QC wave equation, which is the same
for relativistic and non-relativistic systems, was developed in
\cite{SePRA,ClaSo}. In our method, each of the one-dimensional equations
obtained after separation of the QC wave equation is solved by the same
QC method. The QC wave equation (\ref{RelQC}) is separated that gives,
for the potential (\ref{Valr}),
\be
\left[\frac{d^2}{dr^2}+\frac{E^2}4-\left(m-\frac{\tilde\alpha(r)}r
+\sigma r\right)^2-\frac{\vec M^2}{r^2}\right]
\tilde R(r)=0, \label{EqTiR}\ee
\be
\left[\frac{\pa^2}{\pa\theta^2}+\frac 1{\sin^2\theta}
\frac{\pa^2}{\pa\varphi^2}+\vec M^2\right]
\tilde Y(\theta,\varphi)=0. \label{TiYlm}\ee
The angular QC equation (\ref{TiYlm}) determines the squared angular
momentum eigenvalues, $\vec M^2$, which enter into the radial equation
(\ref{EqTiR}). Solution of (\ref{TiYlm}) has been obtained in
\cite{SePRA,An3D} by the QC method in the complex plane that gives
$\vec M^2=(l+\frac 12)^2$. This result means that the radial QC wave
equation has always the centrifugal term $(l+\frac 12)^2/r^2$ for all
spherically symmetrical potentials $V(r)$. The squared angular momentum
eigenvalues, $(l+\frac 12)^2$, are universal for all central potentials
and not any Langer-type corrections are required \cite{SePRA,An3D}.

This QC method reproduces the exact energy eigenvalues for all
known solvable problems in quantum mechanics \cite{SePRA}. In our
QC method not only the total energy, but also momentum of a
particle-wave in bound state is the {\em constant of motion}.
Solution of the QC wave equation in the whole region is written
in elementary functions as \cite{ClaSo},
\be
\tilde R(r)=C_n\left\{\ba{lc} \label{osol}
\frac 1{\sqrt 2}e^{|p_n|r -\phi_1}, & r<r_1,\\
\cos(|p_n|r -\phi_1 -\frac\pi 4), & r_1\le r\le r_2,\\
\frac{(-1)^n}{\sqrt 2}e^{-|p_n|r +\phi_2}, & r>r_2,
\ea\right.\ee
where $C_n=\sqrt{2|p_n|/[\pi(n+\frac 12)+1]}$ is the normalization
coefficient, $p_n$ is the corresponding eigenmomentum,
$\phi_1=-\pi(n+\frac 12)/2$ and $\phi_2=\pi(n+\frac 12)/2$
are the values of the phase-space integral at the turning points
$x_1$ and $x_2$, respectively. In the classically allowed region
[$x_1,x_2$], the solution is
\be
\tilde R_{nl}(r)=C_n\cos\left(|p_n|r+\frac{\pi }2n\right),
\label{TiR}\ee
i.e., has the form of a standing wave. This solution is appropriate
for two-turning-point problems both in non-relativistic and
relativistic cases with the corresponding eigenmomenta $p_n$.

For example, in case of the non-relativistic Coulomb problem,
the total energy eigenvalues have the form of kinetic energy of
a free particle \cite{SePRA,ClaSo},
\be
E_n=\frac{p_n^2}{2m},\ \ \ p_n=\frac{i\alpha m}{n_r+l+1},
\label{Etokin}\ee
where $p_n=mv_n$ is the non-relativistic momentum eigenvalue
with the imaginary discrete velocity, $v_n=i\alpha/(n_r+l+1)$.
This means, for example, that the motion of the electron in a hydrogen
atom is free, but restricted by the ``walls'' of the potential...
This is {\it free finite motion} of a particle-wave in bound state.
One should note, that the QC eigenfunctions (\ref{TiR}) correspond to
the asymptote of the exact solution of the Schr\"odinger equation,
i.e., the principal term of the asymptotic series of the corresponding
exact solution \cite{ClaSo}.

The radial QC equation (\ref{EqTiR}) has four turning points and can not
be solved analytically by standard methods. Let us use the QC method
to solve the equation. The QC quantization condition appropriate
to (\ref{EqTiR}) in the complex plane is \cite{SeEPL,SeRelQC}:
\be
I=\oint_C\sqrt{\frac{E^2}4-\left(m-\frac{\tilde\alpha(r)}r
+\sigma r\right)^2-\frac{(l+\frac 12)^2}{r^2}}dr
=4\pi\left(n_r+\frac 12\right). \label{FasI}\ee
To calculate the phase-space integral (\ref{FasI}) in the complex
plane we chose a contour $C$ enclosing the cuts (therefore, turning
points and zeros of the w.f.) at $r<0$ and $r>0$ between the turning
points $r_1$, $r_2$ and $r_3$, $r_4$, respectively. Outside the contour
$C$, the problem has two singularities, i.e. at $r=0$ and $\infty$.
Using the standard method of stereographic projection, we should
exclude the singularities outside the contour $C$. Excluding these
infinities we have, for the integral (\ref{FasI}), $I=I_0+I_{\infty}$,
where $I_0=-2\pi(l+\frac 12)$ is contribution of the centrifugal term.
The integral $I_{\infty}$ is calculated with the help of the
replacement of variable, i.e., $z=1/r$, that gives
$I_{\infty}=2\pi(E^2/8\sigma+\tilde\alpha_0)$, where
$\tilde\alpha_0=k\alpha_0$ and $\alpha_0$ is the strong coupling in
the NP regime (\ref{alf0}). Here we took into account the asymptotic
properties of the running coupling (\ref{alfr}) and its derivative:
$\alpha_s(r)=0$, $\alpha_s^\prime(r)=0$ at $r\ra 0$. The calculations
result in the squared total energy eigenvalues,
\be
E_n(J)^2=8\tilde\sigma(2n_r+J+3/2-\tilde\alpha_0). \label{En2cl} \ee
Putting in (\ref{En2cl}) $\tilde\alpha_0=0$ we come to well known
relativistic result for the linear potential,
$E_n(J)^2=8\tilde\sigma(2n_r+J+3/2)$.

It is an experimental fact that the dependence $E_n^2(J)$ is linear
for light mesons. However, at present, the best way to reproduce the
experimental masses of light hadrons is to rescale the entire spectrum
assuming that the masses $M_n$ of the mesons are expressed by the
relation \cite{SemCeu}
\be
M_n^2=E_n^2-C^2, \label{Cresc}\ee
where $C$ is a constant energy (free parameter). Relation (\ref{Cresc})
is used to shift the spectra and appears as a means to simulate the
effects of unknown structure approximately. This constant can be
interpreted as a renormalization of the vacuum energy \cite{Bla}.
It has been suggested that the confinement potential has a complex
Lorentz structure, and the relation (\ref{Cresc}) used to shift
the spectra appears as a means to simulate approximately the effects
of this structure. This constant can be connected with the structure
of the QCD vacuum, filled with $J^{PC}=0^{++}$ transverse electric
glueballs which form a negative energy condensate \cite{DonJo}.

The oscillator-type expression (\ref{En2cl}) does not require any
additional free parameter. It contains the needed shift in the form
of the interference term $-8\tilde\alpha_0\tilde\sigma$ of the coulombic
and linear terms of the potential (\ref{Valr}). Formula (\ref{En2cl})
is good to describe spectra of light hadrons, but not heavy quarkonia.

The asymptotic expression (\ref{En2cl}) is defined by the
singularities at $r=0$ and $r\ra\infty$. The leading singularity in
the phase-space integral (\ref{FasI}) at $r\ra\infty$ is given by the
quadratic term $(\sigma r)^2$ originating from the linear part of the
potential (\ref{Valr}), which gives the dominant contribution in
(\ref{FasI}) at infinity. But the term $\propto r^2$ suppresses
a very important contribution of the coulombic interaction at small
and moderate distances $r$.

It is known that heavy $Q\bar Q$ systems can be treated
non-relativistically and, for low states of heavy quarkonia, the main
contribution to the boundary energy comes from OGE term of the
potential (\ref{Valr}), i.e., in the first approximation one can
neglect the confining linear term. A reliable consideration of the
excited $Q\bar Q$ states requires a completely relativistic treatment.

Let us consider the radial equation (\ref{EqTiR}) just for the coulombic
part $V_S(r)$ of the potential (\ref{Valr}). The QC quantization
condition in the complex plane is
\be
I=\oint_C\sqrt{\frac{E^2}4-\left[m-\frac{\tilde\alpha(r)}r\right]^2
-\frac{(l+\frac 12)^2}{r^2}}dr=2\pi\left(n_r+\frac 12\right),
\label{Icou}\ee
and the integral is calculated analogously to the above case. A
contour $C$ encloses the classical turning points $r_1$ and $r_2$
and cut between them. Using the same solution method of stereographic
projection, we obtain, for the integral (\ref{Icou}) outside the
contour $C$ at $r=0$ and $\infty$: $I=I_0+I_{\infty}$, where
$I_0=-2\pi(l+\frac 12)$ and
$I_{\infty}=2\pi(\tilde\alpha_0 m/\sqrt{-E^2/4+m^2}$. This gives:
\be E_n^2=
4m^2\left[-\left(\frac{\tilde\alpha_0}{n_r+l+1}\right)^2+1\right].
\label{En2c} \ee
Again, here we took into account the asymptotic properties of the
strong running coupling (\ref{alfr}) at $r\ra 0$ and $r\ra\infty$.

Thus, we have two exact analytic expressions (\ref{En2cl}) and
(\ref{En2c}) for two asymptotic components of the potential (\ref{Valr})
(coulombic and linear). Now, we can use the same approach as in
\cite{SeEPL,SeZ,SeA}, i.e., derive the interpolating mass formula for
$E^2_n$, which satisfies both of the above constraints: the exact energy
eigenvalues (\ref{En2cl}) and (\ref{En2c}). To derive such a formula
we use the two-point Pad\'e approximant \cite{Bak},
\be [K/N]_f(z)
=\frac{\sum_{i=0}^Ka_iz^i}{\sum_{j=0}^Nb_jz^j}, \label{Pappr} \ee
with $K=3$ and $N=2$. We take $K=3$ and $N=2$ because this is a
simplest choice to satisfy the two asymptotic limits (\ref{En2cl})
and (\ref{En2c}). Simple calculations give the interpolating mass
formula,
\be
M_n^2=4m^2\left[\frac{2\tilde\sigma}{m^2}\left(2n_r+J +\frac 32
-\tilde\alpha_0\right)-\left(\frac{\tilde\alpha_0}{n_r+J+1}
\right)^2+1\right].\label{En2int}\ee
Note, the two exact asymptotic expressions (\ref{En2cl}) and
(\ref{En2c}) for $E^2_n$ have the form of the squared total energy
for two free relativistic particles, i.e., $E_n^2=4(p^2_n+m^2)$.

The mass formula (\ref{En2int}) is good to describe the mass spectra
both light and heavy quarkonia. To demonstrate its efficiency we
calculate the leading state masses of $\rho$ and $\phi$ families
(see tables 1, 2, where masses are in MeV).

\begin{center}
Table 1. The $\rho$-family $J=l+1$ leading states\\
\vspace{3mm}
\begin{tabular}{lllll}
\hline\hline
Meson&$J^{PC}$&$\ \ E_n^{ex}$&$\ \ E_n^{th}$&Parameters in (\ref{En2int})\\
\hline
$\rho\ (1S)$&$1^{--}$&$\ \ 775$&$\ \ 775$&$\Lambda=487$ MeV\\
$a_2(1P)$&$2^{++}$&$\ 1318$&$\ 1319$&$\sigma=0.137$\,GeV$^2$\\
$\rho_3(1D)$&$3^{--}$&$\ 1689$&$\ 1689$&$m_n=144$\,MeV\\
$a_4(1F)$&$4^{++}$&$\ 2001$&$\ 1989$&~\\
$\rho\ (1G)$&$5^{--}$&$ ~ $&$\ 2249$&~\\
$\rho\ (1H)$&$6^{++}$&$ ~ $&$\ 2481$&~\\
$\rho\ (2S)$&$1^{--}$&$\ 1720$&$\ 1683$&~\\
$\rho\ (2P)$&$2^{++}$&$ ~ $&$\ 1986$&~\\
$\rho\ (2D)$&$3^{--}$&$ ~ $&$\ 2247$&~\\
$\rho\ (2F)$&$4^{++}$&$ ~ $&$\ 2480$&~\\
$\rho\ (3S)$&$1^{--}$&$ ~ $&$\ 2245$&~\\
\hline\hline
\end{tabular}
~\\~\\~\\
Table 2. The $\phi$-family $J=l+1$ leading states\\
\vspace{3mm}
\begin{tabular}{lllll}
\hline\hline
Meson&$J^{PC}$&$\ \ E_n^{ex}$&$\ \ E_n^{th}$&Parameters in (\ref{En2int})\\
\hline
$\phi\ (1S)$&$1^{--}$&$\ 1020$&$\ 1019$&$\Lambda=445$\,MeV\\
$f_2(1P)$&$2^{++}$&$\ 1525$&$\ 1525$&$\sigma=0.125$\,GeV$^2$\\
$\phi_3(1D)$&$3^{--}$&$\ 1854$&$\ 1854$&$m_s=414$\,MeV\\
$f_4(1F)$&$4^{++}$&$\ 2018$&$\ 2119$&~\\
$\phi\ (1G)$&$5^{--}$&$ ~ $&$\ 2349$&~\\
$\phi\ (1H)$&$6^{++}$&$ ~ $&$\ 2556$&~\\
$\phi\ (2S)$&$1^{--}$&$\ 1820$&$\ 1820$&~\\
$\phi\ (2P)$&$2^{++}$&$\ 2011$&$\ 2103$&~\\
$\phi\ (2D)$&$3^{--}$&$ ~ $&$\ 2340$&~\\
$\phi\ (2F)$&$4^{++}$&$ ~ $&$\ 2551$&~\\
$\phi\ (3S)$&$1^{--}$&$ ~ $&$\ 2327$&~\\
\hline\hline
\end{tabular}
\end{center}
In this calculations the frozen strong coupling (\ref{alf0}) depends on
the ratio $\mu_g/\Lambda$. From the fit results we found the optimal
value for the constituent gluon mass, $\mu_g=416$ MeV. The QCD dimensional
parameter $\Lambda$ as others is found from the best fit to the available
particle data \cite{Naka}. For light quarks, we take the average effective
mass, $m_n=(m_u+m_d)/2$ MeV. The best fit to the data is achieved for
$m_n=144$ MeV.

The mass formula (\ref{En2int}) is appropriate to calculate the glueball
masses as well. In this case we use the potential (\ref{Valr}) with the
parameters of $gg$ interaction. Calculation results for the gluonium
leading state masses are shown in table 3.
\begin{center}
Table 3. Glueball  $J=l+2$ leading states\\
\vspace{3mm}
\begin{tabular}{lllll}
\hline\hline
Glueball&$J^{PC}$&$\ \ E_n^{ex}$&$\ \ E_n^{th}$&Parameters in (\ref{En2int})\\
\hline
$f_0(1710)$&$2^{++}$&$\  1710$&$\  1710$&$\Lambda=321$ MeV\\
$f_0\ (1P)\ $&$3^{--}$&$~$&$\ 2405$&$\sigma_a=0.329$\,GeV$^2$\\
$f_0\ (1D)\ $&$4^{++}$&$~$&$\ 2921$&$\mu_g=416$\,MeV\\
$f_0\ (1F)\ $&$5^{--}$&$~$&$\ 3350$&~\\
$f_0\ (2S)\ $&$2^{++}$&$~$&$\ 2898$&~\\
$f_0\ (2P)\ $&$3^{--}$&$~$&$\ 3338$&~\\
$f_0\ (2D)\ $&$4^{++}$&$~$&$\ 3720$&~\\
$f_0\ (3S)\ $&$2^{++}$&$~$&$\ 3711$&~\\
$f_0\ (3P)\ $&$3^{--}$&$~$&$\ 4057$&~\\
$f_0\ (3D)\ $&$4^{++}$&$~$&$\ 4374$&~\\
\hline\hline
\end{tabular}
\end{center}
Equation (\ref{En2int}) is an ansatz [as the potential (\ref{Valr})],
which is based on two exact asymptotic expressions (\ref{En2cl}) and
(\ref{En2c}). It allows us to get an {\it analytic} expression for
Regge trajectories in the whole region. Transform (\ref{En2int}) into
the cubic equation for the angular momentum $J(t=E^2)$,
\be
J^3 +a_1(t)J^2 +a_2(t)J +a_3(t)=0, \label{JE3}\ee
where
$a_1(t)=\lambda(t)+2\tilde n$, $a_2(t)=2\tilde n\lambda(t)+\tilde n^2$,
$a_3(t)=\tilde n^2\lambda(t)-\tilde\alpha^2m^2/(2\tilde\sigma)$,
$\tilde n=n_r+1$,
$\lambda(t)=(-t+4m^2)/(8\tilde\sigma)+2n_r+3/2-\tilde\alpha$.
Equation (\ref{JE3}) has three (complex in general case) roots:
$J_1(t)$, $J_2(t)$, and $J_3(t)$. The real part of the first root,
${\rm Re}\,J_1(t)$ (Chew-Frautschi plot), gives the analytic
expression for Regge trajectories,
\be
{\rm Re}\,\alpha(t)=\left\{\ba{lc}
2\sqrt{-p}\cos(\phi/3)-a_1/3,& Q<0;\\
-a_1/3\ (q=0), & Q=0;\\
f_1+f_2-a_1/3,& Q>0,
\label{Ralt}\ea\right.\ee
where
$\phi(t)=\arccos[-q/\sqrt{-p^3}]$, $p(t)=-a_1^2/9+a_2/3$,
$q(t)=a_1^3/27-a_1a_2/6+a_3/2$, $Q(t)=p^3+q^2$,
$f_1(t)=[-q+\sqrt{Q}]^{1/3}$, $f_2(t)=[-q-\sqrt{Q}]^{1/3}$.
Expression (\ref{Ralt}) supports the existing experimental data and
gives the saturating Regge trajectories including the Pomeron
in the whole region of $t$.

The imaginary part of the complex Regge trajectories is given
by the expression
\be
{\rm Im}\,\alpha(t)=\sqrt 3(f_1-f_2)/2,\ Q>0.
\label{Ialt}\ee
The threshold (trajectory termination point), beyond which no bound
states should exist, is defined from the equation, $Q(t)=p^3+q^2=0$
(see notations above).

\begin{figure}[htb]
\begin{center}
\includegraphics[scale=2]{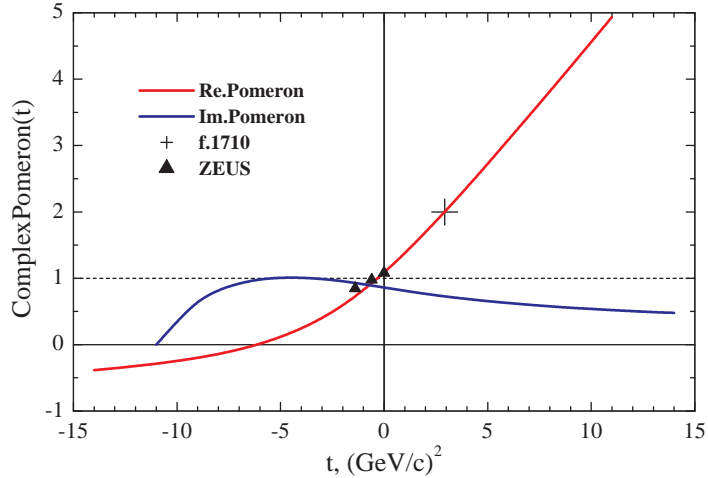}
\caption{{\small The real and imaginary components of the gluonium
$J=l+2$ complex Regge trajectory (the Pomeron). Triangles show combined
HERA $\rho$ and $\phi$ data \protect\cite{ZEUS,Ast,HERA}.}}
\label{Fig3}\end{center}
\end{figure}

The effective $P$ trajectory in figs.\,\ref{Fig1}-\ref{Fig3} has
similar properties as all quark-antiquark trajectories given by
(\ref{Ralt}). It is asymptotically linear at $t\ra\infty$ with the
slope $\alpha_P^\prime=1/(8\sigma_a)\simeq 0.380$\,(GeV/c)$^{-2}$,
and flattens off at $-1$ for $t\ra-\infty$. As all Regge trajectories
the $P$ trajectory is the monotonically rising function at the interval
($-\infty,\infty$).

Parameters of the function Re\,$\alpha_P(t)$ are found from the best
fit to the combined ZEUS $\rho$ (triangles) and $\phi$ (circles)
scattering data \cite{ZEUS}, and glueball candidate $f_0(1710)$
\cite{Kirk} (cross) with quantum numbers $J^{PC}=2^{++}$
($l=0$, $S_z=+2$) in bound state region. The intercept and slope of
the $P$ trajectory at $t=0$ are:
\be
\alpha_P^\prime(0)=1.083,\ \ \ \alpha_P^\prime(0)=0.280 {\rm (GeV/c)}^2.
\label{Ppar}\ee
The hard BFKL Pomeron has intercept $\alpha_P(0)\simeq 1.43$ \cite{BFKL}.
The Pomeron with such properties results in too fast growth of the total
cross sections.
Recent studies of exclusive electro-production of vector mesons at JLab
\cite{CLAS,CLAS1} made it possible for the first time to play with two
independent hard scales: the virtuality $Q^2$ of the photon, which sets
the observation scale, and the momentum transfer $t$ to the hadronic
system, which sets the interaction scale. They reinforce the description
of hard scattering processes in terms of few effective degrees of
freedom relevant to the Jlab-Hermes energy range \cite{Lag04}.

The study of exclusive electro-production of $\omega$ mesons,
completed at JLab \cite{CLAS,CLAS1}, provides us with an original
insight on the space time structure of hard scattering processes
between the constituents of hadrons. It was shown that the higher order
mechanisms are more economically described in terms of a few effective
degrees of freedom: dressed parton propagators, saturating Regge
trajectories and electromagnetic form factors of off-shell meson.
The success of this description in several channels is a strong hint
that they are the relevant degrees of freedom in the JLab-Hermes energy
range. In addition, they provide us with a link with more fundamental
approaches of NP QCD: {\it ab initio} Lattice Gauge calculations or
potential models.

The two-Pomeron picture (soft plus BFKL Pomeron) gives a very good fit
to the total cross section for elastic $J/\Psi$ photoproduction and
the charm structure function $F_2^c$ over the whole range of $Q^2=-t$
\cite{CDL}. The hard BFKL Pomeron has the intercept
$\alpha_{{\rm BFKL}}(0)\simeq 1.43$. Next-to-leading order estimates
give, for the BFKL intercept values $1.26$ to $1.30$, which is closer
to the soft supercritical Pomeron. However, the results of the
experiments and the found higher order corrections make it quite unclear
what the BFKL Pomeron is. Another question is: what is the intercept
of the BFKL Pomeron, if the pQCD is non-applicable in this scale, i.e.,
at $t=0$\,? On the other hand, the saturated $P$ trajectory (\ref{Ralt})
has the properties of the soft supercritical Pomeron at small $-t$ and
saturates at large $-t$ according to the pQCD prediction.

\section{Conclusion}
We have considered glueballs as bound states of constituent massive
gluons and investigated their properties in the framework of the
potential approach. The constituent gluon picture could be questioned
since potential models have serious difficulties in reproducing all
the currently known lattice QCD data. In spite of non-relativistic
phenomenological nature, the potential approach can be used to describe
glueballs. Fair description of quarkonium states give us a confidence
that we are on the right way to describe glueballs.

The physical properties of constituent gluons are still a matter of
controversy. Within the framework of potential models, gluons are
supposed to be massless or massive, i.e., with either a helicity-1 or
a spin-1 particles. We have dealt with simplest two-gluon glueballs,
but mass and spin in different works are very different. If valence
gluons are assumed to be helicity-1 particles, then their spin has
only two projections: $S_z=\pm 1$. In this work gluons are considered
as massive spin-1 particles with projections $S_z=-1,0,+1$. The
properties of these pure glue states are not completely understood.

We have analyzed two exact asymptotic solutions of relativistic QC
wave equation for the QCD-inspired scalar potential with the coordinate
dependent strong coupling, $\alpha_s(r)$. One needs to stress that the
behavior of the potential and exact form of the coupling $\alpha_s(r)$
are not so important in the intermediate region: in fact, the results
depend on the asymptotic behavior of the potential at $r\ra 0$ and
$r\ra\infty$. Our QC method in the complex plane to resolve the
eigenvalue problem is rather simple and allows to get the analytic
result. Using two asymptotes, corresponding to the short-distance
coulombic and long-distance linear components of the potential, we
have derived the universal mass formula (\ref{En2int}) and calculated
the glueball masses, which are in agreement with the lattice data.

We have considered glueballs as the physical particles on the $P$
trajectory. To reproduce the trajectory, we have inverted the mass
formula (\ref{En2int}) and derived the analytic expression (\ref{Ralt})
for the $P$ trajectory, ${\rm Re}\,\alpha_P(t)$, and its imaginary
part ${\rm Im}\,\alpha_P(t)$ in the whole region of $t$. In the
scattering region, at $-t\gg\Lambda_{QCD}$, the trajectory flattens
off at $-1$, i.e., it has asymptote $\alpha_P(t\ra-\infty)=-1$
(saturates). In bound state region, at large timelike $t$, the $P$
trajectory is linear in accordance with the string model.

It is known, that the fixed-number of particles with a potential
description can not be used for strict relativistic description.
Strict description of the Pomeron presupposes multiparticle
description of the system. For perturbative regime with the
Pomeron scattering, the dominant contribution comes from the BFKL
Pomeron. However, experimental data and our simple calculations in
the framework of the potential approach support the conception of
the soft supercritical Pomeron as observed at the presently
available energies. The saturating Regge trajectories (\ref{Ralt})
obtained in this work effectively include short- and long-distance
dynamics of constituents.

In this paper we have not considered helicities and spin
properties of gluons. This topic has been discussed in details
somewhere else \cite{MaBuSe}. The existing data and simple
analysis performed in this work confirm the existence of the
Pomeron which is the complex non-linear function with the
properties of soft supercritical Pomeron at small $-t$ and
saturates Pomeron at large squared momentum transfers.

\section*{Acknowledgement}
The author would like to thank N.M. Shumeiko, A.A. Pankov and
Yu.A. Kurochkin for support and constant interest to this work.

\end{document}